\documentclass{elsart}

\usepackage{graphics}
\usepackage{epsfig}
\usepackage{amssymb}

\begin{document}

\begin{frontmatter}

\title{Modern theories of low-energy astrophysical reactions}

\author[pisa]{L.\ E.\ Marcucci},
\author[anl]{Kenneth M.\ Nollett},
\author[jlabodu]{R.\ Schiavilla}, and
\author[anl]{R.\ B.\ Wiringa}
\address[pisa]{Department of Physics, University of Pisa, and INFN-Sezione di 
Pisa, I-56127 Pisa, Italy}
\address[anl]{Physics Division, Argonne National Laboratory, Argonne, 
Illinois 60439}
\address[jlabodu]{Jefferson Lab, Newport News, Virginia 23606 \\
         and
         Department of Physics, Old Dominion University,
         Norfolk, Virginia 23529}

\begin{abstract}

We summarize recent {\it ab initio} studies of low-energy electroweak
reactions of astrophysical interest, 
relevant for both big bang nucleosynthesis and solar neutrino
production.  The calculational methods include direct integration for
$np$ radiative and $pp$ weak capture, correlated hyperspherical
harmonics for reactions of $A=3,4$ nuclei, and variational Monte Carlo
for $A=6,7$ nuclei.  Realistic nucleon-nucleon and three-nucleon
interactions and consistent current operators are used as input.
\end{abstract}

\end{frontmatter}

\section{Realistic interactions and opportunities for astrophysics}
\label{sec:intro}

\def\dalphg{$d(\alpha,\gamma)^6{\rm Li}$}
\def\tag{$^3{\rm H}(\alpha,\gamma)^7{\rm Li}$}
\def\hag{$^3{\rm He}(\alpha,\gamma)^7{\rm Be}$}
\def\bep{$^7{\rm Be}(p,\gamma)^8{\rm B}$}

Combined partial-wave analyses \cite{nijmphase} of essentially all
elastic nucleon-nucleon ($N\!N$) scattering data have been used to
construct highly accurate potential models
\cite{wss95,machleidt94,nijmII} that describe the $N\!N$
interaction with high precision, the ``realistic'' interactions.
Available methods allow nuclear wave functions, binding energies, and
electroweak matrix elements to be computed from these potentials for
nuclei containing up to ten nucleons \cite{ppwc01,pvw} with great
success in matching the laboratory data.  All of this is done with
only bare interactions between pairs or triples of nucleons, and not
with the effective interactions tailored to each class of problem that
are more typical of nuclear physics.  In particular, simpler potential
models \cite{dubovichenko95,kim81,buck85,buck88,Mohr93} of radiative
captures in light nuclei need spectroscopic factors provided by
experiment, while the resonating group and similar techniques
\cite{kim81,walliser83,kajino84,walliser84,liu81,fujiwara83,kajino86,mertelmeier86,altmeyer88,csoto00}
rely on nuclear interactions tailored to each system and use limited
model spaces that affect the accuracy of cross section calculations in
difficult-to-predict ways \cite{altmeyer88,csoto00}.

The mass and energy range accessible at present to calculations from
realistic interactions and currents corresponds to two important areas
of astrophysics: nucleosynthesis during the big bang
\cite{Wag77} and hydrogen burning by the $pp$ chain in low-mass stars
including the sun \cite{BC38,adelberger}.  The present paper reviews the
methods and results of recent theoretical work on several rates
important for these problems.

The chemical composition of the universe just after the big bang was
set by the freezeout of nuclear reactions when the universe was less
than about five minutes old, a process called big-bang nucleosynthesis
(BBN) \cite{Wag77,kolbturner}.  The main products are hydrogen and
\nuc{4}{He}, in a ratio set mainly by weak interactions when the
universe was about one second old.  Small amounts of \nuc{2}{H},
\nuc{3}{He} and \nuc{7}{Li} were also made, and their abundances after
BBN provide crucial information on the mean baryon density in the
universe \cite{schrammturner}.  Theoretical predictions of these
abundances depend on the cross sections for eleven key nuclear
reactions \cite{skm}, all empirically determined quantities.
At present, the reactions
accessible to the methods discussed here are the radiative captures
$d(p,\gamma)^3\mathrm{He}$, $p(n,\gamma)d$, \hag, and \tag.  The first
two of these are crucial for the \nuc{2}{H} yield of BBN, and the last three
are crucial for the \nuc{7}{Li} yield.

The rates of energy and neutrino generation in the sun are set by the
rate of the first reaction in the $pp$ chain, the weak capture $p+p
\longrightarrow d+e^++\nu_e$ \cite{BC38}.  The cross section for this
process is so small that it cannot be measured in the laboratory, so
it must be provided by theory.  The rates of other $pp$-chain
reactions (reviewed by Haxton, Rolfs and Parker in this volume) do not
appreciably affect energy generation in the Sun, but they do set the
flux and energy distribution of solar neutrinos.  In particular, the
radiative captures \hag\ and \bep\ are accessible in the laboratory,
but the Coulomb barrier and the low ($\sim 20$ keV) energies involved
make the cross sections very small and the experiments difficult.  The
absolute normalization is the crucial property of a cross section for
astrophysics, and this is precisely what is most difficult to measure
reliably in the laboratory.  A theoretical treatment must be grounded
in first principles in order to provide independent information about
absolute cross sections.  Theoretical calculations are also possible
for weak decays of \nuc{7}{Be} and \nuc{8}{B} that actually produce
the neutrinos.  An additional process important for the solar neutrino
spectrum, the weak capture ${^3}\mathrm{He} + p \longrightarrow\
^4\mathrm{He} + e^++\nu_e$ (the ``$hep$'' reaction), is accessible to
these methods but not at all to experiment.

\section{The nuclear Hamiltonians}
\label{sec:nucham}
The Hamiltonians we use include nonrelativistic one-body kinetic energy,
any of several modern, accurate two-nucleon potentials, and a three-nucleon
potential motivated by meson-exchange or chiral effective field theory:
\begin{equation}
   H = \sum_{i} K_{i} + \sum_{i<j} v_{ij} + \sum_{i<j<k} V_{ijk} \ .
\end{equation}
The kinetic energy operator is predominantly charge-independent (CI),
but has a small charge-symmetry-breaking (CSB) component due to the
difference between proton and neutron masses.  The modern
high-precision two-nucleon ($NN$) potentials 
we consider are the Argonne $v_{18}$ (AV18)
\cite{wss95}, CD-Bonn \cite{machleidt94}, Nijm-I, Nijm-II, and Reid93
\cite{nijmII}.  All these potentials are charge-dependent (CD),
reproducing both the $pp$ and $np$ scattering lengths, as well as the
deuteron binding energy, and fitting a total of 4301 $N\!N$ scattering
data from the 1993 Nijmegen partial wave analysis (PWA93)
\cite{nijmphase} up to $E_{lab}=350$ MeV with a $\chi^2/$datum
$\approx 1$.  

The AV18 potential is given by a sum of electromagnetic, one-pion-exchange (OPE),
and short-range phenomenological terms.  
The strong interaction part is written as a sum of operator components:
\begin{eqnarray}
\label{eq:operator}
   v_{ij} &=& \sum_{p=1,18} v_{p}(r_{ij}) O^{p}_{ij} \, \\
   O^{p=1,8}_{ij} &=& [1, {\bf\sigma}_{i}\cdot{\bf\sigma}_{j}, S_{ij},
   {\bf L\cdot S}]\otimes[1,{\bf\tau}_{i}\cdot{\bf\tau}_{j}] \ .
\end{eqnarray}
The first eight operators are the most important, reproducing the CI average
of S- and P-wave phase shifts, while
six additional terms quadratic in momentum are required to fit higher
partial waves, and four small CD and CSB terms are required to explain
fine differences between $pp$, $np$, and $nn$ scattering.
AV18 is local in every partial wave, and all partial waves are connected 
by this underlying operator structure.

All the other potentials are defined partial wave by partial wave.
The Nijm-II and Reid93 potentials are local in each partial wave,
while the Nijm-I central force includes momentum-dependent terms that
give rise to non-local structures in configuration space.  The CD-Bonn
potentials have non-local components in both central and tensor parts
of the interaction.  All these modern potentials have about 40
parameters adjusted to fit the $N\!N$ data.  They cover a range of 
possible behaviors, as exhibited in their deuteron wave functions, 
shown in Fig.~1 of Ref.~\cite{Sch98}.  Despite these differences in 
detail, all five models give almost identical deuteron observables.

The realistic $N\!N$ potentials give varying amounts of binding in the
triton, ranging from 7.6 MeV for the local potentials to 8.0 MeV for
the most non-local, compared to the experimental value of 8.48 MeV.  In 
the $\alpha$-particle, the binding ranges from 24.0 to 26.3 MeV, compared 
to the experimental 28.3 MeV \cite{Nog00}.  Thus empirically, we need to
add some three-nucleon ($3N$) potential to obtain the proper binding
and size of light nuclei.  On the basis of meson-exchange
theory we certainly expect $3N$ potentials to play a role, while
chiral perturbation theory suggests there should be a rapid convergence
in the contributions of many-body forces, so we may neglect
four-nucleon forces.

The $3N$ forces we use are the Tucson-Melbourne (TM) potential
\cite{TM79}, its chirally-improved successor TM$^\prime$ \cite{fhk99},
or one of the Urbana potentials \cite{CPW83}.  These take the general form:
\begin{equation}
   V_{ijk} = V^{2\pi,S}_{ijk} + V^{2\pi,P}_{ijk} + V^{R}_{ijk} \ .
\label{eq:tni}
\end{equation}
The TM and TM$^\prime$ potentials are constructed from two-pion exchange,
where an intermediate $\pi\!N$ scattering takes place in either S- or P-waves.
The pion-range functions embedded in these terms have a cutoff that
may be adjusted to fit the triton binding energy when used in
conjunction with a particular $N\!N$ potential.  The Urbana potentials
neglect the small S-wave term, but add a phenomenological
short-range repulsion, $V^{R}_{ijk}$.  In this case, the pion-range
functions are taken to be the same as in the $N\!N$ potential,
while the strength of the terms are adjusted to
reproduce the binding energy of the triton and to give a reasonable
saturation density in nuclear matter.  Urbana model IX (UIX) \cite{PPCW95}
was adjusted specifically to go with AV18.
A more sophisticated set of Illinois model $3N$ potentials has been
constructed recently \cite{ppwc01}, which include an additional three-pion
ring term, $V^{3\pi,\Delta R}_{ijk}$.  The combined AV18/IL2 Hamiltonian
reproduces light nuclear binding energies very well up to $A=10$, but it
has not yet been used in astrophysical rate calculations.

\section{Transition rates}
\label{sec:transition}
The capture processes of interest in the present study involve
radiative or weak transitions between an initial two-cluster continuum
state $| {\bf p}; J_1M_1,J_2 M_2\rangle^{(+)}$, with clusters 
$A_1$ and $A_2$ having spins $J_1 M_1$ and $J_2 M_2$, respectively, 
and relative momentum ${\bf p}$, and a final $A$-nucleon 
bound state $|J_AM_A\rangle$.
The cross section and polarization observables are obtained from
matrix elements of the current operator connecting these two states.
For example, the spin-averaged differential cross section for 
radiative capture in the center-of-mass (CM) reference 
frame is written as

\begin{eqnarray}
\label{eq:jme}
\frac{d\sigma^\gamma}{d\Omega} (E,\theta) &
=&\frac{\alpha}{2\pi\, v} \frac{q}{1+q/m_A}\\
&&\times\sum_{\lambda M_A} \overline{\sum_{M_1 M_2}}
\mid 
\langle -{\bf q};J_A M_A \mid \hat{\bf \epsilon}_\lambda^*({\bf q}) \cdot
{\bf j}^\dagger({\bf q})\mid p \hat{\bf z}; J_1M_1,J_2 M_2\rangle^{(+)}
\mid^2 \ ,
\nonumber
\end{eqnarray}
where ${\bf q}$ is the momentum of the emitted photon, $\hat{\bf
\epsilon}_\lambda({\bf q})$, $\lambda=\pm 1$, are the spherical
components of its polarization vector, and ${\bf j}({\bf q})$ is the
nuclear electromagnetic current operator (see Sec.~\ref{sec:nto}
below).  The final bound state is recoiling with momentum $-{\bf q}$.
The $\hat{\bf p}$-direction has been taken to define the spin
quantization axis, and $p$ depends on the relative energy
$E$=$p^2/(2\mu)$, $\mu$ being the $A_1$-$A_2$ reduced mass.  The angle
$\theta$ is the angle between $\hat {\bf z}$ and $\hat {\bf q}$,
$\alpha$ is the fine-structure constant, $m_A$ is the bound-state
mass, $v=p/\mu$ is the relative velocity, and the photon energy $q$ is
fixed by energy conservation.  The dependence of the observables upon
$\theta$ can be derived from the expansion 
of the initial capture state $| {\bf p}; J_1M_1,J_2 M_2\rangle^{(+)}$ 
into partial waves $|LSJM\rangle^{(+)}$, 
and from the multipole expansion of the current operator~\cite{Car98}.  
The total cross section, however, is simply given by

\begin{eqnarray}
\sigma^\gamma(E)&=& \frac{2\,\alpha}{v} \frac{q}{1+q/m_A}  
\frac{1}{(2J_1+1)(2J_2+1)} \nonumber \\
&\times& \sum_{l \geq 1}\sum_{LSJ} \Big[\mid\! E_l(q;LSJ)\!\mid^2
             +\mid\! M_l(q;LSJ)\!\mid^2\Big] \ ,
\end{eqnarray}
where $E_l$ and $M_l$ denote the reduced matrix elements of the electric
and magnetic multipoles, i.e.  $E_l(q;LSJ) \equiv \langle J_A\mid\mid
E_l(q)\mid\mid LSJ\rangle^{(+)}$ and similarly for $M_l(q;LSJ)$.

The weak processes considered in the present study are proton weak 
capture on $^1$H and $^3$He, nuclear $\beta^\pm$-decay and \nuc{7}Be 
decay by electron ($\epsilon$-) capture. The cross section for the 
proton weak capture reactions is written as

\begin{equation}
\sigma^\beta(E)=\int 2\pi \, \delta(E_i-E_f)\frac{1}{v}
\sum_{s_e s_\nu}\overline{\sum_{M_1M_2}}
|\langle f\,|\,H_W\,|\,i\rangle|^{2}
\frac{d{\bf{p}}_{e}}{(2\pi)^3} \frac{d{\bf{p}}_{\nu}}{(2\pi)^3} \ ,
\label{eq:xscb}
\end{equation}
where $H_W$ is the weak-interaction Hamiltonian~\cite{Wal95}, $E_i$
and $E_f$ are the initial and final state energies (including rest
masses), and ${\bf p}_e$ and ${\bf p}_\nu$ are the positron/electron
and electron-neutrino momenta.  The rates for the $\beta^{\pm}$ decays
and $\epsilon$-capture are given by a similar expression as above, but
with the flux factor $1/v$ removed and, in the case of the
$\epsilon$-capture, the phase-space integration involves only the
outgoing neutrino momenta.

It is convenient to write the avereged weak-interaction Hamiltonian 
matrix element as $\sum_{s_e s_\nu} \overline{\sum_{M_1 M_2}}
|\langle f\,|\,H_{W}\,|\,i\rangle|^{2} 
= (2 \pi)^2\> G_V^2\> L_{\sigma \tau} \> N^{\sigma\tau}$,
where $G_V$ is the Fermi coupling constant ($G_V$=1.14939$\times 10^{-5}$
GeV$^{-2}$), and the lepton tensor $L^{\sigma \tau}$ can
be expressed in terms of the positron/electron  
and neutrino four velocities~\cite{Mar01}.  
The distortion of the charged-lepton wave function in the Coulomb 
nuclear field is also considered in $L^{\sigma \tau}$ 
for processes with $A>4$~\cite{SW02}.
The nuclear tensor $N^{\sigma \tau}$ can be expressed in terms of the 
reduced matrix elements of the Coulomb, longitudinal, electric
and magnetic transition operators, corresponding to the multipole
expansion of the nuclear weak current.  Since the latter has scalar or
vector $(V)$ and pseudoscalar or axial-vector $(A)$ components, each
multipole consists of the sum of $V$ and $A$ terms, 
and the parity of $l$th multipole
$V$-operators is opposite of that of $l$th multipole $A$-operators.
The parity of Coulomb, longitudinal, and electric $l$th multipole
$V$-operators is $(-)^l$, while that of magnetic $l$th multipole
$V$-operators is $(-)^{l+1}$.

Finally, in later sections, the results for capture reactions between
charged clusters will be presented in terms of the astrophysical
$S$-factor, related to the total cross section $\sigma(E)$, via
$S(E)=E\sigma(E) {\rm exp}(2\pi\, Z_1 Z_2
\alpha/v)$, where $Z_1$ and $Z_2$ are the atomic numbers of the
incoming clusters.  The exponential factor is the Coulomb penetration
factor.

\section{The nuclear transition current operators}
\label{sec:nto}
The model for the nuclear electroweak current has been most recently
reviewed in Refs.~\cite{Car98,Mar01}. The current is written as sum 
of one- and many-body components that operate on the nucleon degrees 
of freedom.
The one-body operators are derived from the nonrelativistic reduction
of the covariant single-nucleon electroweak currents and include
terms up to order $1/m^2$, where $m$ is the nucleon
mass~\cite{Car98,Mar01}.  Inclusion of only one-body terms is known as
the ``impulse approximation'' (IA).  The two-body
operators are discussed briefly in the following two
subsections.

\subsection{The electromagnetic two-body currents}
\label{subsec:emc}

The two-body current has ``model-independent'' and ``model-dependent''
components, in the classification scheme of Riska~\cite{Ris89}.  The
model-independent terms are obtained from the two-nucleon interaction
and are constructed~\cite{Ris85} so as to satisfy current conservation.
The leading operator is the isovector ``$\pi$-like'' current obtained
from the isospin-dependent spin-spin and tensor
interactions~\cite{Car98}.  The same components of the
interaction also generate an isovector ``$\rho$-like'' current.
Additional model-independent isoscalar and isovector currents arise
from the isospin-independent and isospin-dependent momentum-dependent interactions.
However, these currents are short-ranged and numerically far less
important than the $\pi$-like current.
 
The model-dependent currents are purely transverse and therefore
cannot be directly linked to the two-nucleon interaction.  Present
results include the contributions due to the isoscalar
$\rho\pi\gamma$ and isovector $\omega\pi\gamma$ transition
currents as well as those due to the isovector current associated with
excitation of intermediate $\Delta$ resonances.  Two different
approximations have been used to treat the $\Delta$ degrees of
freedom~\cite{Car98}; one is based on first-order perturbation theory,
using the static $\Delta$ approximation, while the other is the
transition-correlation-operator (TCO) scheme~\cite{Sch92}, essentially
a scaled-down version of the full $N$+$\Delta$ coupled-channel method.
Comparisons between results obtained within these two approximate
schemes have been reported in a number of
studies~\cite{Sch92,Viv96,Mar98}.

The presence of $3N$ interactions in the Hamiltonian requires
corresponding three-body currents in order to fulfill current
conservation, but these were found to be numerically unimportant in
studies of the trinucleon magnetic form factors at intermediate
momentum transfer \cite{Mar98}.

\subsection{The weak two-body currents}
\label{subsec:wkc}

The nuclear weak current consists of vector ($V$) and axial-vector ($A$) parts,
with corresponding one- and two-body components.  The weak vector
current is constructed from the isovector part of the electromagnetic
current, in accordance with the
conserved-vector-current hypothesis \cite{Wal95}.

Some of the two-body axial-current operators are derived from $\pi$-
and $\rho$-meson exchanges and the $\rho\pi$-transition mechanism.
These mesonic operators have been found to give rather small
contributions in weak transitions involving few-nucleon
systems~\cite{Sch98,Mar01,Mar02}.  The two-body weak axial-charge
operator includes a pion-range term from soft-pion theorem and current
algebra arguments~\cite{Kub78}, as well as short-range terms
associated with scalar- and vector-meson exchanges.  The latter are
obtained consistently with the two-nucleon interaction model,
following a procedure~\cite{Kir92} similar to that used to derive the
corresponding weak vector-current operators~\cite{Mar01}.
 
The dominant two-body axial current operator, however, is that due to
$\Delta$ excitation~\cite{Sch98,Mar01}.  Since the $N\Delta$
transition axial-vector coupling constant $g_A^*$ is not known
experimentally, the associated contribution suffers from a large model
dependence.  To reduce it, $g_A^*$ has been
adjusted to reproduce the experimental value of the Gamow-Teller (GT)
matrix element in \nuc{3}{H} $\beta$ decay (cf. Table \ref{tab:weak}).
As in the case of the electromagnetic current, the perturbative and
TCO approximations have been used to calculate the $\Delta$-excitation
axial current contributions.  We emphasize that results obtained
within the two schemes are typically very close to each other once
$g_A^*$ is fixed, for each scheme, to reproduce the \nuc{3}{H} GT
matrix element (cf. Table~XV of Ref.~\cite{Mar01}).

Some of our weak transition studies in few-nucleon systems 
have also been carried out by using weak current
operators derived within an effective-field-theory (EFT) approach, in
which pions and nucleons are the explicit degrees of freedom and terms
up to next-to-next-to-next-to-leading order (N$^3$LO) are included.
Explicit expressions for these EFT operators can be found in
Ref.~\cite{Par03}.  Here we only note that: i) the two-body operators are
regularized by a cutoff that defines the energy/momentum scale of EFT 
below which the chosen explicit degrees of freedom are valid; 
ii) the axial-current terms depend on an
unknown parameter $d_R$ that gives the strength of a counterterm and 
is fixed, for a given value of the
cutoff, by again reproducing the \nuc{3}{H} GT matrix element.
The premise of EFT is that
physical observables should not depend on the cutoff as long as the
latter is chosen in a physically reasonable range.  Such an expectation
is indeed borne out in the EFT calculation of the $pp$ capture
discussed below.

\section{Wave functions}
\label{sec:wf}

Wave functions (bound and continuum) for the present calculations
are computed by two means: correlated hyperspherical harmonics (CHH)
and variational Monte Carlo (VMC).
In the CHH method, the nuclear wave function is expanded on a suitable
basis and the unknown expansion parameters are calculated using
appropriate variational principles. This technique is very successful
in describing nuclear ground states with $A\leq
4$~\cite{Kie93,Viv95,Nog03,Kam01}, the $N\!d$ scattering states both below and
above the deuteron breakup threshold~\cite{Kie94,Kie95,Kie99,Kie00,Kie01},
and the $p$\nuc{3}He and $n$\nuc{3}H systems below the \nuc{3}He and
\nuc{3}H breakup thresholds~\cite{Viv98,Viv01}.
VMC is an approximate method that uses Monte Carlo sampling to perform
numerical quadratures and obtain upper bounds to ground and excited
state energies for $3\leq A\leq 10$ nuclei~\cite{pvw,Pie01}.  
There is also the Green's function Monte Carlo (GFMC) method that starts 
from a VMC trial function and employs Monte Carlo integration to evaluate 
an imaginary-time path integral that projects out exact states of a nucleus.
Extension of the Monte Carlo methods to describe continuum states is
more difficult, as discussed below, and the GFMC method has not yet been
used to compute reaction cross sections.

\subsection{Hyperspherical harmonic wave functions for bound states}
\label{sec:boundchh}

We first consider the CHH method applied to the $A=3$ bound state.
The wave function $\Psi$ of a three-nucleon system with total angular 
momentum $JM$ and total isospin $TT_{z}$ can be written 
as~\cite{Kie93}
\begin{equation}
\Psi=\sum_{\alpha,n}u_{\alpha,n}(\rho)
     \sum_{p} H_{\alpha,n}(p) \ .
\label{eq:psi3}
\end{equation}
Here $\rho$ is the hyper-radius, $\rho=(x_i^2+y_i^2)^{1/2}$, with the 
Jacobi coordinates ${\bf x}_i$ and ${\bf y}_i$ defined respectively as 
${\bf x}_i={\bf r}_j-{\bf r}_k$ and ${\bf y}_i= 
\left({\bf r}_j+ {\bf r}_k-2\,{\bf r}_i\right)/{\sqrt{3}}$. 
The index $p$ denotes an even permutation of the particle indices 
$i$,$j$,$k$.
The functions $H_{\alpha,n}$ are antisymmetric under the
exchange $j\leftrightharpoons k$. They are given by the product of an
angle-spin-isospin function $Y_\alpha$ and a hyperspherical
polynomial $P_{\alpha,n}(\phi_i)$, $\phi_i$ being the hyper-angle
defined as $\phi_i=\cos^{-1}(x_i/\rho)$.  The index $\alpha$ specifies
the set of the spectator $i$ and pair $jk$ orbital angular momenta,
$l_\alpha$ and $L_\alpha$, spin, and isospin, which are
coupled to produce the given $JMTT_{z}$ quantum numbers
and the appropriate parity.  The $P_{\alpha,n}(\phi_i)$ is proportional 
to a Jacobi polynomial; the second index $n$ runs over all
non-negative integers and specifies the order $K=l_\alpha+L_\alpha+2n$
of the polynomial.  The product of the Jacobi polynomial with
the spherical harmonics $Y_{l_\alpha}(\widehat{\bf x}_i)$ and
$Y_{L_\alpha}(\widehat {\bf y}_i)$ contained in $Y_\alpha$ is
by definition a hyperspherical harmonic function.  Finally,
correlation factors, which are functions of the $jk$ relative
distance, are included in $H_{\alpha,n}$ to account for the
strong state-dependent correlations induced by the $N\!N$
interaction. These correlation functions are solutions of suitable
two-body Schr\"odinger equations~\cite{Kie93}.
Their presence greatly improves the rate of convergence of the
expansion of Eq.~(\ref{eq:psi3}); the results quoted in this work 
have been obtained with 23 angle-spin-isospin channels, 
for which the maximum $K$ is 16~\cite{Nog03}.  
To obtain a comparable accuracy without correlation factors, 
the maximum value of $K$ needs to be about 180~\cite{Kie97}.

The Rayleigh-Ritz variational principle is used to determine the
hyper-radial functions $u_{\alpha,n}(\rho)$ and the ground state
energy $E$. Carrying out the variation $\delta_u\Psi$ with respect to
the functions $u_{\alpha,n}(\rho)$, and performing the spin-isospin
sums and the integration over the angular and hyper-angular variables
yields a set of coupled second-order differential equations for the
$u_{\alpha,n}(\rho)$.  This is then solved
numerically~\cite{Kie93,Kie94}.

The extension of the CHH approach to the study of the
$\alpha$-particle is conceptually straightforward but numerically 
much more involved~\cite{Viv95,Viv98}.  
In analogy with the three-body case, the wave function $\Psi$ is 
written as in Eq.~(\ref{eq:psi3}) and the Rayleigh-Ritz
variational principle is again used to determine the hyper-radial
functions and ground state energy $E$.
However, it should be noted that in the case of $A=4$, there are 
two sets of Jacobi coordinates, corresponding to the 1+3 and 2+2 
partitions, and both partitions have been 
used in the definition of the angle-spin-isospin functions $Y_\alpha$. 
Furthermore, there are three hyperspherical coordinates, the hyper-radius 
and two hyper-angles, and only one hyper-angle depends on the partition 
considered. The hyperspherical polynomial of Eq.~(\ref{eq:psi3}) depends 
in this case on two non-negative integers, which specify the order of two
individual Jacobi polynomials, functions of the hyper-angles.
Finally, the correlation factors are products of pair-correlation 
functions and are obtained by the same procedure as in the three-body 
case.  More details are given in Ref.~\cite{Viv95}.  

The results for the binding energies of $^3{\rm H}$, $^3{\rm He}$ and
$^4{\rm He}$ obtained with the CHH method for the AV18/UIX Hamiltonian
are given in Table~\ref{tab:be}~\cite{Mar01,Nog03}.  Note that
an accuracy of 1 (10) keV can be achieved for the three-body
(four-body) problem.

\begin{table}
\caption{Calculated AV18/UIX and experimental binding energies
of light nuclei in MeV.}
\begin{tabular}{llllll}
\hline
$^AZ(J^{\pi};T)$
        &   CHH     &   VMC(I)  &  VMC(II)  &   GFMC     &  Expt  \\
\hline
$^3$H$(\frac{1}{2}^+;\frac{1}{2})$
        & ~8.479    & ~8.227(6) &           & ~8.461(6)  & ~8.482 \\
$^3$He$(\frac{1}{2}^+;\frac{1}{2})$
        & ~7.750    & ~7.476(6) &           & ~7.708(6)  & ~7.718 \\
$^4$He$(0^+;0)$     
        & 27.89     & 27.40(3)  &           & 28.31(2)   & 28.30  \\
$^6$Li$(1^+;0)$     
        &           & 28.05(5)  & 28.16(5)  & 31.25(8)   & 31.99  \\
$^7$Li$(\frac{3}{2}^-;\frac{1}{2})$
        &           & 33.07(7)  & 32.47(7)  & 37.5(1)    & 39.24  \\
$^7$Li$(\frac{1}{2}^-;\frac{1}{2})$
        &           & 33.13(7)  & 32.23(7)  & 37.6(1)    & 38.76  \\
$^7$Be$(\frac{3}{2}^-;\frac{1}{2})$
        &           & 31.49(7)  & 30.60(7)  & 35.9(1)    & 37.60  \\
$^7$Be$(\frac{1}{2}^-;\frac{1}{2})$
        &           & 31.58(8)  & 30.71(7)  & 35.9(2)    & 37.17  \\
\hline
\end{tabular}
\label{tab:be}
\end{table}

\subsection{Variational Monte Carlo wave functions for bound states}
\label{sec:boundvmc}

A good VMC wave function, $\Psi_V(J^\pi;A,T)$, is constructed from 
products of two- and three-body correlation operators acting on an 
antisymmetric Jastrow wave function with the appropriate quantum numbers:
\begin{equation}
     |\Psi_V\rangle = \left[1 + \sum_{i<j<k\leq A} U^{TNI}_{ijk} \right]
                      \left[ {\mathcal S}\prod_{i<j\leq A}(1+U_{ij}) \right]
                      |\Psi_J\rangle \ .
\label{eq:psit}
\end{equation}
Here $\mathcal S$ denotes symmetrization over the noncommuting pair 
correlation operators $U_{ij}$, which include spin, isospin, and 
tensor terms, $ U_{ij} = \sum_{p=2,6} u_p(r_{ij}) O^p_{ij}$, and 
$O^{p}_{ij}$ are the same static operators that appear in
the $N\!N$ potential, Eq.(\ref{eq:operator}). The $U^{TNI}_{ijk}$ 
are correlations induced by the three-nucleon interaction, and have
the same spin-isospin structure as the $V_{ijk}$ of Eq.(\ref{eq:tni}).

The Jastrow wave function, $\Psi_J$, for s-shell nuclei has
the simple form
\begin{equation}
     |\Psi_J\rangle = \left[ \prod_{i<j<k\leq A}f^c_{ijk} \right]
                      \left[ \prod_{i<j\leq A}f_c(r_{ij}) \right]
                     |\Phi_A(JMTT_{z})\rangle \ .
\label{eq:jastrow}
\end{equation}
Here $f_c(r_{ij})$ and $f^c_{ijk}$ are central two- and three-body correlation
functions and $\Phi_A$ is a Slater determinant in spin-isospin space, e.g.,
for the $\alpha$-particle, $|\Phi_{4}(0 0 0 0) \rangle
= {\mathcal A} |p\uparrow p\downarrow n\uparrow n\downarrow \rangle$.
The functions $f_c(r)$ and $u_p(r)$ are generated by solving a set 
of coupled differential equations containing the bare $NN$ potential
with appropriate asymptotic boundary conditions~\cite{W91}.  Variational parameters
in these equations are adjusted to minimize the energy expectation value, 
$E_V = \langle \Psi_V | H | \Psi_V \rangle$, which is evaluated by 
Metropolis Monte Carlo integration~\cite{Met53}.

The $\Psi_J$ for p-shell nuclei is more complicated, with single-nucleon 
p-shell orbitals and $LS$ coupling to obtain the desired $JM$ value 
\cite{CK65}.  We sum over all allowed spatial symmetries $[n]$
\cite{BoMo69} and allow for three types of central pair correlations
depending on which shell the nucleons are in:
\begin{eqnarray}
  |\Psi_{J}\rangle &=& {\mathcal A} \left\{\right.
     \prod_{i<j<k\leq 4} f^{c}_{ijk} \prod_{i<j \leq 4} f_{ss}(r_{ij})
     \sum_{LS[n]} \Big( \beta_{LS[n]}
     \prod_{k \leq 4<l \leq A} f^{LS[n]}_{sp}(r_{kl})           \nonumber\\
  && \prod_{4<l<m \leq A} f^{LS[n]}_{pp}(r_{lm})
    |\Phi_{A}(LS[n]JMTT_{z})_{1234:5\ldots A}\rangle \Big) \left.\right\} \ .
\label{eq:jastrow-p}
\end{eqnarray}
The operator ${\mathcal A}$ indicates an antisymmetric sum over all 
possible partitions into 4 s-shell and $(A-4)$ p-shell particles.
The pair correlation for both particles within the s-shell, $f_{ss}$,
is similar to the $f_c$ of the $\alpha$-particle.
The pair correlations for both particles in the p-shell, $f^{LS[n]}_{pp}$,
and for mixed pairs, $f^{LS[n]}_{sp}$, are similar to $f_{ss}$ at short
distance, but their long-range structure is adjusted to give appropriate 
clustering behavior, and they may vary with $LS[n]$.
The single-particle wave function components are:
\begin{eqnarray}
 &&  |\Phi_{A}(LS[n]JMTT_{z})_{1234:5\ldots A}\rangle =
     |\Phi_{4}(0 0 0 0)_{1234}\rangle \times 
     |\prod_{4 < l\leq A} \phi^{LS[n]}_{p}(R_{\alpha l}) \\
 &&  \left\{ [ \prod_{4 < l\leq A} Y_{1m_l}(\Omega_{\alpha l}) ]_{LM_L[n]}
     \times [ \prod_{4 < l\leq A} \chi_{l}(\frac{1}{2}m_s) ]_{SM_S}
     \right\}_{JM}
     \times [ \prod_{4 < l\leq A} \nu_{l}(\frac{1}{2}m_t) ]_{TT_z}\rangle
     \nonumber \ .
\end{eqnarray}
The $\phi^{LS[n]}_{p}(R_{\alpha l})$ are p-wave solutions of a particle
in an effective $\alpha$-$N$ potential that has Woods-Saxon and Coulomb parts.
They are functions of the distance between the center of mass
of the $\alpha$-core and nucleon $l$, and may vary with $LS[n]$.
The wave function is translationally invariant, so there is no
spurious center of mass motion.

Two different types of $\Psi_J$ have been constructed in recent VMC
calculations of light p-shell nuclei: Type I, which is a shell-model
trial function~\cite{PPCPW97}, and Type II, which is a cluster-cluster
trial function~\cite{NWS01,Nol01}.  In Type I trial functions, the
$\phi^{LS[n]}_{p}(r)$ has an exponential decay at long range, with the
depth, range, and surface thickness of the Woods-Saxon potential
serving as variational parameters.  The $f^{LS[n]}_{sp}$ go to a
constant near unity at long range, while the $f^{LS[n]}_{pp}$ have
a small long-range tail that is larger for states of lesser spatial
symmetry $[n]$ \cite{PPCPW97}.
In Type II trial functions, $\phi^{LS[n]}_{p}(r)$ is again the solution
of a p-wave Schr\"odinger equation with a potential containing
Woods-Saxon and Coulomb terms, but with an added Lagrange multiplier
that turns on at long range and imposes the boundary condition
$[\phi^{LS[n]}_{p}(r\rightarrow\infty)]^n \propto W_{km}(2\gamma r)/r$.
Here $W_{km}(2\gamma r)$ is the Whittaker function for bound-state wave
functions in a Coulomb potential and $n$ is the number of p-shell nucleons.
The cluster separation wave number $\gamma$ is taken from experiment.
The accompanying $f^{LS[n]}_{sp}$ goes to unity 
while the $f^{LS[n]}_{pp}$ are 
taken from the exact deuteron wave function in the case of \nuc{6}{Li}, or
the VMC \nuc{3}{H} (\nuc{3}{He}) $\Psi_V$ in the case of \nuc{7}{Li}
(\nuc{7}{Be}).
Consequently, Type II trial functions factorize at large cluster
separations as $\Psi_V \rightarrow \psi_{\alpha} \psi_\tau
                W_{km}(2\gamma r_{\alpha\tau})/r_{\alpha\tau}$,
where $\psi_{\alpha}$ and $\psi_\tau$ are the wave functions of the
clusters and $r_{\alpha\tau}$ is the separation between them
\cite{NWS01,Nol01}.  

For either type of trial function, a diagonalization is carried out in
the one-body basis to find the optimal values of the $\beta_{LS[n]}$
mixing parameters for a given $(J^\pi;T)$ state.
Current best VMC energies for nuclear states of astrophysical interest are
given in Table~\ref{tab:be} for the AV18/UIX Hamiltonian along with CHH
and GFMC results and experimental values.  These VMC results are about 1
MeV more bound for $A$=6,7 nuclei than previously reported because of
recent improvements in the parametrization of the trial functions;
however, results for various transition matrix elements are not
significantly altered.  The VMC energies for $A$=3,4 nuclei
are 1--2\% above CHH or GFMC results, but are 10--15\% higher for
$A$=6,7 than GFMC results, which are themselves 5\% high compared to
experiment with this Hamiltonian.

\subsection{Hyperspherical harmonic wave functions for continuum states}
\label{sec:freechh}

The CHH method has also been applied to the study of scattering
problems. In particular, $pd$ and $nd$ systems have been studied both
below~\cite{Kie94,Kie95} and above~\cite{Kie99,Kie00,Kie01} the
deuteron breakup threshold.  The $p\,^3{\rm He}$ and $n\,^3{\rm H}$
systems have also been investigated below the $^3{\rm He}$ ($^3{\rm
H}$) breakup threshold~\cite{Viv98,Viv01}.  We now review how the CHH
method works for the $N+A$ scattering below the mass-$A$ nuclear
breakup threshold.

The wave function $\Psi_{N+A}^{LSJM}$, having incoming orbital angular
momentum $L$ and spin $S$ coupled to total $JM$, is written as the sum
of a core function $\Psi_{c}^{JM}$ and an asymptotic function
$\Psi_{a}^{LSJM}$.  The function $\Psi_c$ describes the system in the
region where the particles are close to each other and their mutual
interactions are large, and vanishes in the limit of large
inter-cluster separation. It is obtained by an expansion on the same
CHH basis as discussed above for the bound state.  In the asymptotic
region the inter-cluster interactions are negligible, and
$\Psi_a^{LSJM}$ is written as (in the $p+A$ case)
\begin{eqnarray}
\Psi_{a}^{LSJM}   &=& \sum_{i} \sum_{L^\prime S^\prime}
    \Big[ \lbrack s_i \otimes \phi_A
    \rbrack_{S^\prime} \otimes Y_{L^\prime}({\hat {\bf r}}_{pA}) \Big]_{JM}
   \nonumber \\
   &&\times \Bigg\lbrack \delta_{L L^\prime} \delta_{S S^\prime}
   { F_{L^\prime}(kr_{pA}) \over kr_{pA} }
    + R^J_{LS,L^\prime S^\prime}(k)
   { G_{L^\prime}(kr_{pA}) \over kr_{pA} }
   g (r_{pA}) \Bigg\rbrack
   \>\>\>. \label{eq:psia}
\end{eqnarray}
Here $\phi_A$, ${\bf r}_{pA}$ and $k$ are the mass-$A$ cluster wave
function, the proton and $A$-cluster relative distance, and the
magnitude of the relative momentum, respectively.  The functions $F_L$
and $G_L$ are the regular and irregular Coulomb functions,
respectively.  For $n+A$ scattering, $F_L(x)/x$ and $G_L(x)/x$ are
replaced by the regular and irregular spherical Bessel functions.  The
function $g(r_{pA})$ regularizes $G_L(kr_{pA})$ at $r_{pA}=0$, and
$g(r_{pA})\rightarrow 1$ as $r_{pA} \geq 10-12$ fm.  Finally
$R^J_{LS,L^\prime S^\prime}(k)$ are the $R$-matrix elements that
determine phase shifts and (for coupled channels) mixing angles at the
energy $k^2/(2\mu)$, $\mu$ being the $N+A$ reduced mass.  The sum over
$L^\prime S^\prime$ includes all values compatible with a given total
$J$ and parity.  Both the matrix elements $R^J_{LS,L^\prime
S^\prime}(k)$ and the hyper-radial functions occurring in the
expansion of $\Psi_c$ are determined by applying the Kohn variational
principle.

The $nd$, $pd$, and $p\,^3{\rm He}$ scattering lengths predicted by
the AV18/UIX model are shown in Table~\ref{tb:sl}~\cite{Viv96,Viv98}.
Also listed in the table are the $n\,^3{\rm H}$ scattering lengths
predicted by the older AV14/UVIII Hamiltonian~\cite{W91,WSA84}.  There
is excellent agreement between the theoretical predictions and the
available data.

\begin{table}
\caption{Calculated $A=3$ doublet ($a_d$) and quartet ($a_q$), and $A=4$ 
singlet ($a_s$) and triplet ($a_t$) scattering lengths in fm~\cite{Viv96,Viv98}.
Experimental results are from Refs.~\cite{Dil71,All93}.}
\label{tb:sl} 
\begin{tabular}{l c c c c c c c c }
\hline 
 & \multicolumn{2}{c}{$nd$} & \multicolumn{2}{c}{$pd$} 
 & \multicolumn{2}{c}{$n\,^3{\rm H}$} & \multicolumn{2}{c}{$p\,^3{\rm He}$} \\
 & $a_d$ & $a_q$ & $a_d$ & $a_q$ 
 & $a_s$ & $a_t$ & $a_s$ & $a_t$ \\
\hline
CHH  & 0.63 & 6.33 & -0.02 & 13.7 & 4.32 & 3.80 & 11.5 & 9.13 \\
Expt & 0.65(4) & 6.35(2) & & & & & 10.8(2.6) & 8.1(5) \\
\hline
\end{tabular}
\end{table}

\subsection{Phenomenological continuum states for $A>4$}
\label{sec:freevmc}
Few calculations of continuum states consisting of two nuclei have
been done in the quantum Monte Carlo formalism, and several
difficulties in these calculations remain to be addressed.  We expect
these problems to be tractable, and work on them continues.  In the
meantime, we have adopted a semi-phenomenological prescription for the
continuum-state wave functions that enter into capture calculations
for the $A>4$ systems \cite{NWS01,Nol01}.

At large separations between nuclei in a continuum state, the wave
function is the product of the wave functions $\phi_1$ and $\phi_2$ of
the two nuclei with a two-body correlation that describes the (nuclear
plus Coulomb) scattering of the two nuclei from each other.  This is
\begin{equation}
\Psi^{LSJM}_{A_1+A_2}  \propto
    \Big[ \lbrack \phi_1  \otimes \phi_2
    \rbrack_{S} \otimes Y_{L}({\hat {\bf r}}_{12}) \Big]_{JM}
   \nonumber 
   \Bigg\lbrack 
   { F_{L}(kr_{12}) \over kr_{12} }
    + \tan \delta_{LSJ}(k)
   { G_{L}(kr_{12}) \over kr_{12} }
   \Bigg\rbrack,
\label{eqn:vmcasymp}
\end{equation}
where $r_{12}$ is the separation between the initial nuclei.  This is
analogous to Eq.(\ref{eq:psia}) without coupled channels, and the
angular momentum labels have the same meanings.

To obtain approximate continuum wave functions, we have adopted the
form
\begin{equation}
\Psi^{LSJM}_{A_1+A_2}  = \mathcal{A} \psi_{12}^{LSJ}(r_{12}) 
    \prod_{ij} G_{ij}
    \Big[ \lbrack \phi_1  \otimes \phi_2
    \rbrack_{S} \otimes Y_{L}({\hat {\bf r}}_{12}) \Big]_{JM}
   \nonumber 
\label{eqn:freevmc}
\end{equation}
for the wave function.  The operator $\mathcal{A}$ antisymmetrizes the
wave function under exchange of nucleons between the two (internally
antisymmetric) clusters.  The short-range $G_{ij}$ operator
correlations between nucleon pairs in different clusters produces
distortions in the nuclei when they occupy the same space, and they go
to unity at less than 2 fm interparticle separation.  We used
correlations $G_{ij}$ found in nuclear-matter calculations
\cite{lagaris81} and found that they had essentially no effect on the
cross sections.

We generated the radial function $\psi_{12}^{LSJ}(r_{12})$ from a
Schr\"odinger equation
\begin{equation}
 \left\{ \frac{\hbar^2}{2\mu_{12}} 
 \left[ -\frac{d^2}{d r_{12}^2} + \frac{L(L+1)}{r_{12}^2} \right] 
      + V_{LSJ}(r_{12}) \right\} (r_{12} \psi_{12}^{LSJ})
  = E (r_{12} \psi_{12}^{LSJ}).
\label{eqn:freeschroedinger}
\end{equation}
The appropriate reduced mass for the nuclei is $\mu_{12}$.  The
effective potential function $V_{LSJ}(r_{12})$ was chosen to be the
sum of a short-range nuclear interaction (Woods-Saxon or similar, with
central and spin-orbit terms) and the Coulomb interaction.  We chose
$V_{LSJ}(r_{12})$ to reproduce the experimentally determined phase
shifts $\delta_{LSJ}(k)$.  For both $A=6$ and $A=7$ $\alpha$-captures,
several suitable choices of $V_{LSJ}$ were already available in the
literature, so we chose from among those
\cite{dubovichenko95,kim81,buck85,buck88,langanke86,kp1,kp2}, making
minor adjustments to reproduce the well-known width and location of
the $3^+$ resonance in \nuc{6}{Li}.  The dependence of our results on
the different $V_{LSJ}$ indicates the amount of uncertainty introduced
by this approach.  To satisfy the Pauli principle, the functions
$V_{LSJ}$ need deep potential wells with forbidden states to produce
the correct number of nodes in $\psi_{12}^{LSJ}$ \cite{zaikin71}, as
well as parity dependence.  The \nuc{6}{Li} $V_{LSJ}$ also contains a
spin-orbit term to describe three low-lying resonances important for
the radiative capture.  It can in principle also admit a tensor
interaction (requiring modification of Eqs.(\ref{eqn:vmcasymp}) and
(\ref{eqn:freeschroedinger})), but it is not unambiguously required by
the data, and we did not include it.

The $J^\pi=1^+$ wave functions in $\alpha d$ should also be to be
orthogonal to the \nuc{6}{Li} ground state, but there is no guarantee
of this in the present approach because $\psi_{12}^{LSJ}$ was not
obtained from the underlying $NN$ interaction.  We found that this
property is only important for magnetic dipole transitions proceeding
from $S$ waves.  For this matrix element, we kept only the part of
$\Psi^{LSJM}_{A_1+A_2}$ orthogonal to the VMC ground state.

Finally, the states $\phi_1$ and $\phi_2$ appearing in
Eq.(\ref{eqn:freevmc}) were computed directly from the two- and
three-nucleon forces by the VMC method, except that the deuteron wave
function $\phi_d$ was a direct numerical solution of the two-body
Schr\"odinger equation for the deuteron.

\section{Results}
\label{sec:results}

\subsection{The $p$($n$,$\gamma$)$d$ radiative capture below 1 MeV}
\label{sec:a2r}

Historically, the radiative capture of thermal neutrons on protons
played a crucial role in establishing the quantitative importance of
two-body current effects in photonuclear observables~\cite{Ris72}.
Their inclusion resolved the long-standing discrepancy between the
calculated impulse approximation (IA) cross section and the measured
value.  In this section, we briefly summarize the results of cross
section calculations for this process from thermal energies up to
energies of $\simeq$ 1 MeV; the energy range between a few tens of keV
up to 200 keV is particularly relevant for primordial nucleosynthesis.

At thermal energy, the cross section $\sigma^\gamma$ is due to the
well-known isovector magnetic dipole ($M_1$) transition connecting the
$^1S_0$ $np$ scattering state to the deuteron bound state.  There is
in principle an isoscalar $M_1$ transition proceeding through the
$^3S_1$ $np$ continuum state, but the associated contribution (at
thermal energy) is strongly suppressed in IA by orthogonality;
furthermore, isoscalar two-body currents, such as those originating from 
the momentum-dependent components of the two-nucleon interaction or from
the $\rho\pi\gamma$ transition mechanism, are numerically far less
important than isovector currents, and play a very marginal role in
this transition.

The calculated values for $\sigma^\gamma$ are listed in
Table~\ref{tb:npdg}, both for one-body currents alone and for the one-
and two-body currents.  The IA results are to a large extent
determined by the $^1S_0$ scattering length and deuteron binding
energy, which the above interactions reproduce exactly.  Indeed, in
effective range theory, one finds that $\sigma^\gamma \simeq 300$ mb.
Among the two-body current contributions, which collectively amount to
approximately 9\% of the total cross section, the largest one, about
60\%, comes from the currents associated with pion exchange.  The next
to largest, about 25\%, is from currents involving $\Delta$
excitation, that have been treated in perturbation theory here.

\begin{table}
\caption{Total cross-section $\sigma^\gamma$ in mb for thermal $np$ radiative 
capture; data from Ref.~\protect\cite{Mugh81}.}
\begin{tabular}{lcccc}
\hline
$NN$ model  & AV18  & Nijm-I & CD-Bonn & Expt     \\
\hline
One-body    & 304.6 & 305.4  & 306.5   &          \\
Full        & 334.2 & 332.5  & 331.6   & 332.6(7) \\
\hline
\end{tabular}
\label{tb:npdg}
\end{table} 
 
As the energy increases from a few to several hundreds of keV,
contributions to $\sigma^\gamma$ from the electric dipole ($E_1$)
transitions proceeding through the $^3P_J$ $np$ scattering states
become more important, and in fact dominant at energies larger than
300 keV.  The Siegert form is used for the $E_1$ operator, and
relativistic corrections to the charge density, such as those
associated with spin-orbit and pion-exchange
contributions~\cite{Viv01a,NWS01}, have been neglected.

\begin{figure}[bth]
\centerline{\epsfig{file=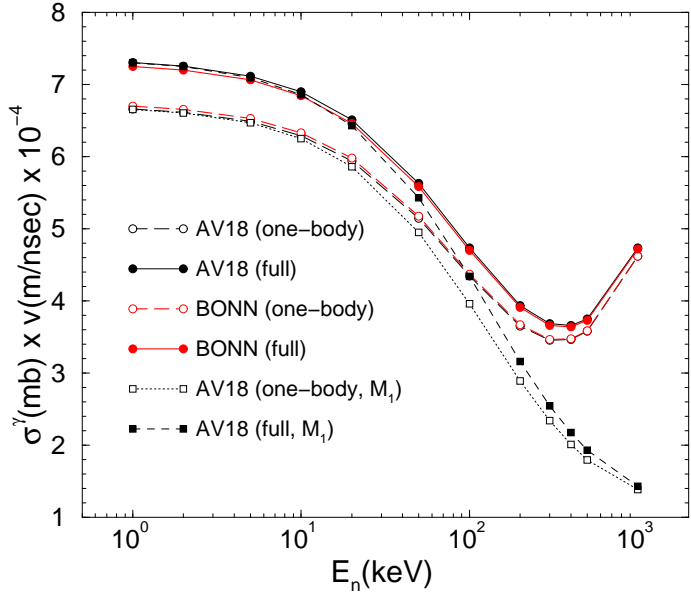,height=8cm}} 
\caption{Predictions for $np$ radiative capture cross section, calculated
with the AV18 and CD-Bonn interactions.  
Also shown for AV18 is the $M_1$ contribution to $\sigma^\gamma$.}
\label{fig:xnpdg}
\end{figure}

The results of calculations based on the AV18 and CD-Bonn interactions
are shown in Fig.~\ref{fig:xnpdg}.  As the
energy increases the model dependence, already less than 1\% at
thermal energies, becomes negligible at 1 MeV, where the
$E_1$ contribution dominates.  This is to be expected, since the $E_1$
transition is mostly sensitive to the long-range part of the wave
functions.
Fig.~\ref{fig:xnpdg} demonstrates that the model dependence of the
predictions in the conventional approach based on realistic
interactions is considerably less than 1\% in the energy range
relevant for primordial nucleosynthesis.  It could be further
reduced by adjusting, for example, the transition
magnetic moment $\mu_{\gamma N\Delta}$ of the $\Delta$-excitation current to precisely
fit the thermal neutron capture cross section.
In all results reported so far $\mu_{\gamma N\Delta}$ has been taken
to be 3 nm, a value consistent with an analysis of $\gamma N$ data at resonance.

Finally, it is worth emphasizing that the model dependence referred to
above is comparable to the theoretical uncertainty inherent in
calculations based on effective field theory~\cite{Che99,Rup00}.
These calculations involve contact operators whose strength has to be
constrained by experimental data.  In the more recent study of
Ref.~\cite{Rup00}, for example, there is a contact four-nucleon-photon
operator, whose $M_1$ ($E_1$) strength is fixed by the measured
thermal capture cross section (deuteron photo-disintegration data at
threshold).

\subsection{The $pp$ weak capture}
\label{sec:a2w}

The theoretical description of the proton weak capture on protons was
first given by Bethe and Critchfield~\cite{BC38}, who showed that the
associated rate was large enough to account for the energy released by
the sun.  Since then, a series of calculations have refined their
original estimate either by computing the required wave functions more
accurately~\cite{Sal52,BM69,GG90,Kam94} or by using more realistic
models for the nuclear transition operator~\cite{GH72,DRR76,CRSW91}.
However, the most recent studies~\cite{Sch98,Par03} offer an
integrated treatment of these two aspects with emphasis on a reliable
estimate of their associated uncertainties.

The $pp$ weak capture is induced by the weak axial current 
operator, and the $S$-factor is conventionally written in terms of a
dimensionless parameter $\Lambda(E)$~\cite{Sch98}, where $E$ is the
$pp$ relative energy.  When axial two-body currents are neglected, it
is given by

\begin{equation}
   \Lambda(E) = \left(\gamma^3/2\right)^{1/2}
                \frac{ 1 }{C_0(k) \, k}
                \int_0^{\infty}dr\,u(r) \chi_0(r;k) \ ,  \label{LAMPRACT}
\end{equation}
where $\gamma$ is the deuteron wave number, $k$ is the $pp$ relative
momentum ($E$=$k^2/m_p$), and $C_0(k)$ is the Gamow penetration
factor.  In the overlap integral, $u(r)$ is the S-wave component of
the deuteron wave function and $\chi_0(r;k)$ is the $pp$ $^1S_0$
continuum wave function.  The $^1S_0$ wave function is the solution of
a Schr\"odinger equation with a $pp$ interaction including, in
addition to the nuclear and Coulomb terms, terms of second order in
the fine structure constant.  These are the vacuum polarization and
two-photon exchange terms, as well as corrections originating from the
finite size of the proton~\cite{Sch98}.  Properly accounting for these
additional electromagnetic interaction components leads to some
practical problems in integrating the Schr\"odinger equation, matching
the Coulomb functions, and extracting the phase shift.  The most
important correction to the Coulomb interaction between protons is the
vacuum polarization, which reduces the cross section (proportional to
$\Lambda^2$) by about 1\%.  Other fine details of the electromagnetic
interaction increase the cross section by about 0.1\%.  This is in
part canceled by a net 0.03\% reduction in cross section from the
correct relativistic treatment of the deuteron wave number, $\gamma$.
Including just the axial one-body operator, five modern interaction
models differ by only 0.3\% in the calculated cross section as
illustrated in Table~\ref{tb:lmb}.

\begin{table}
\caption{Square of the overlap integral $\Lambda(E=0)$
for five modern $N\!N$ interaction models.}
\begin{tabular}{lccccc}
\hline
$NN$ model  & AV18 & Nijm-I & Nijm-II & Reid93 & CD-Bonn \\
\hline
$\Lambda^2$ (one-body) & 6.965 & 6.965 & 6.971 & 6.974 & 6.985 \\
$\Lambda^2$ (full)     & 7.076 &       &       &       & 7.060 \\
\hline
\end{tabular}
\label{tb:lmb}
\end{table}
 
The model for the axial two-body current contains the simplest
possible two-body operators that give an adequate description of the
longest-range mechanisms.  The overall strength of the leading
operator due to $\Delta$ excitation is adjusted as described in
Sec. \ref{subsec:wkc}.  The contributions due to exchanges of heavier
mesons, such as the $A_1$~\cite{Cie84}, or renormalization effects
arising from $\Delta$ admixtures in the nuclear wave
functions~\cite{Sch92}, are neglected.  However, in Ref.~\cite{Sch98}
it was shown that these approximations do not influence in any
significant way the theoretical predictions for the $pp$ weak capture
cross section once the two-body current model is constrained to fit
the GT matrix element of tritium.

The results for $\Lambda^2(0)$ including axial two-body currents are
listed for the AV18 and CD-Bonn interactions in Table~\ref{tb:lmb}.
Predictions for this quantity with other modern interactions are
expected to be similar.  Thus, the model dependence and theoretical
uncertainty appear to be at the level of a few parts in a thousand,
much smaller than the estimate given in Ref.~\cite{adelberger}.

More recently~\cite{Par03}, the $pp$ (and $hep$, see below) 
astrophysical rates have been
calculated using wave functions obtained from
solutions of the Schr\"odinger equation with realistic
interactions, and axial two-body currents derived from chiral
effective field theory (EFT), as discussed at the end of 
Sec.~\ref{subsec:wkc}. In Ref.~\cite{Par03}, the $pp$ $S$-factor is 
the physical observable chosen to test the energy/momentum 
cutoff dependence. It has been demonstrated that this observable 
is independent of the cutoff as long 
as its value is within a physically reasonable range.   
In particular, this work gives for the ratio of
the two-body to one-body contributions the value $(0.86 \pm 0.05)$\%,
in line with the results of the conventional theory reported above.

\subsection{The $d(n,\gamma)$\nuc{3}H and 
$d(p,\gamma)$\nuc{3}He radiative captures}
\label{sec:dndp}

The theoretical description of the $A$=3 and 4 radiative and weak
capture processes constitutes a challenging problem for nuclear
few-body theory~\cite{Mar01,Car98,Sch92,Car90,Mar00}.  The difficulty
comes about because the $A$=3 and 4 bound states are approximate 
eigenstates of the $M_1$ and GT one-body operators~\cite{Sch37}.  
This property would be exact if these states were to consist only of a 
symmetric S-wave component, which accounts for over 90\% (80\%) of 
the $A$=3 (4) bound states.  It is spoiled by the presence of D-state 
components induced by tensor interactions.  Consequently,
the one-body matrix elements of these (largely) $M_1$- or GT-induced
processes are suppressed due to orthogonality between the initial and
final states.  As a result, the associated cross sections become
extremely sensitive to contributions from small components in the wave
functions and two-body electroweak current operators.

The extensive experimental data for the $d(n,\gamma)$\nuc{3}H and 
$d(p,\gamma)$\nuc{3}He reactions include both total cross sections and spin
polarization observables at several center-of-mass energies.  
The most complete comparison between theory and experiment
for the $d(n,\gamma)$\nuc{3}H and $d(p,\gamma)$\nuc{3}He radiative
captures below the deuteron breakup threshold was performed using CHH
wave functions obtained from the AV18/UIX interaction~\cite{Viv96,Viv01a},
with a nuclear electromagnetic current operator that includes 
both one- and two-body contributions.  The proton radiative capture 
calculation has been extended above the deuteron breakup threshold 
\cite{Viv03,Mar03a,Mar03b}, but we consider only the low-energy regime.

The theoretical prediction of Ref.~\cite{Viv96} for the total cross
section $\sigma_T$ of thermal $d(n,\gamma)^3\mathrm{H}$ is
0.578 for the full and 0.229 mb for the 
impulse approximation (IA) current operators.  The
experimental value is $\sigma_T = 0.508\pm0.015$ mb~\cite{Jur82}.
Therefore, the ``full'' result, while clearly an improvement over the
IA, still exceeds the experimental value by 14\%.
The origin of this discrepancy is puzzling, particularly in
view of the fact that the astrophysical $S$-factor for the related $dp$ radiative
capture at zero energy is calculated to be within 1\% of the measured value
(see below).

\begin{figure}
\centerline{\epsfig{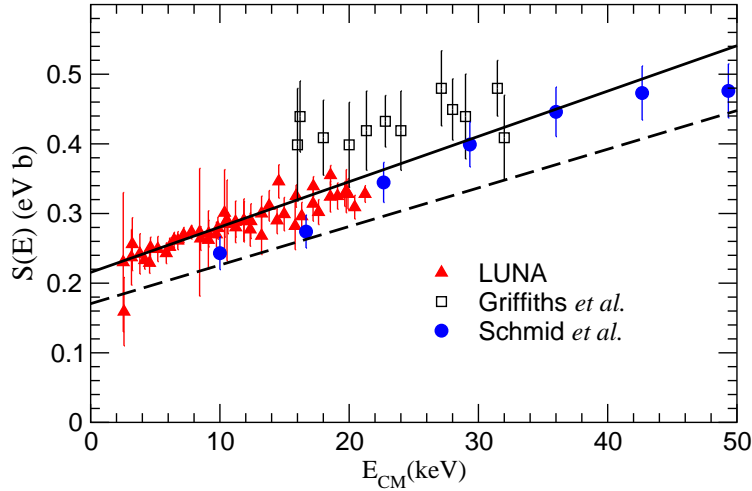}}
\caption{The calculated $S$-factor~\cite{Viv01a} for 
$d(p,\gamma)^3\mathrm{He}$ obtained using AV18/UIX
CHH wave functions and one-body only (dashed line) or both one- and
two-body (solid line) currents.  Experimental data are taken from
Refs.~\cite{LUN02,Gri63,Sch95a,Sch95b}.\label{fig:dpcapture}}
\end{figure}

The calculated $S$-factor~\cite{Viv01a} in the energy range 0--50 keV
is compared with the experimental data of
Refs.~\cite{LUN02,Gri63,Sch95a,Sch95b} in Fig.~\ref{fig:dpcapture}.
The solid curve represents the ``full'' result, while the
dashed line is the result in IA.  The agreement between theoretical
predictions and experimental data, especially the very recent LUNA data
\cite{LUN02}, is excellent.  The two-body contributions to the nuclear
electromagnetic current operator play a very important role in this
agreement.  The ``full'' (IA) theoretical value for $S(0)$ has been
found to be 0.219 (0.162) eV$\cdot$b, in excellent agreement with the
LUNA result of $0.216\pm 0.010$ eV$\cdot$b.  The theoretical and
experimental $S$-factors also agree nicely at higher energies, as
shown in Refs.~\cite{Viv03,Mar03b} in the energy range 0--2 and
0--3.33 MeV.

\subsection{The $hen$ and $hep$ reactions}
\label{sec:henhep}

The total cross section of the \nuc{3}He$(n,\gamma)$\nuc{4}He ($hen$)
reaction is experimentally known~\cite{Wer91}, while the astrophysical
$S$-factor of the \nuc{3}He$(p,e^+\,\nu_e)$\nuc{4}He $(hep)$ reaction
cannot be directly measured in the energy range relevant for solar fusion.
The most recent {\em ab initio} calculation of the $hen$ reaction was
performed more than a decade ago in Refs.~\cite{Sch92,Car90}, using
VMC wave functions obtained with the older AV14/UVIII Hamiltonian
model.  The total cross section at thermal neutron energy was found to
be 85.9 $\mu$b when one- and two-body contributions to the
electromagnetic current are included, and 5.65 $\mu$b in IA.  The most
recent experimental determination gives 55$\pm$3
$\mu$b~\cite{Wer91}. The discrepancy between theory and experiment
could indicate that: i) the VMC wave functions obtained with the
AV14/UVIII interaction model are not sufficiently accurate; and/or ii)
the model for the electromagnetic current needs to be refined.  Work
currently under way to construct more accurate CHH wave functions for the
four-nucleon states involved in the $hen$ reaction will help to clarify 
this issue.

The Super-Kamiokande (SK) measurements of the energy spectrum of
electrons recoiling from scattering with solar
neutrinos~\cite{SK01,Fuk99,Suz00,Fuk02} have brought new interest in
the $hep$ reaction.  The first SK results for
this spectrum showed an apparent excess of events in the
highest-energy bin~\cite{Fuk99} that could be explained by increasing
by a factor of $\simeq$ 17 the Standard Solar Model (SSM98)
prediction~\cite{BBP98} of the $hep$ neutrino flux.  The most recent
results from SK are essentially the same after taking the large mixing
angle (LMA) neutrino mixing parameters into account.  However, when
the quasi-vacuum solution is used, the inferred flux at the earth is
reduced by a factor of 4.  The $hep$ neutrino flux is directly
proportional to the $S$-factor at zero energy $S(0)$, and the SSM98
estimate for the flux was based on the calculation of
Ref.~\cite{Sch92}.  This calculation used AV14/UVIII VMC wave
functions, retained only the $p$\nuc{3}He $^3S_1$ channel, neglected
the dependence on the momentum transfer of the lepton pair, and used
poorly-constrained two-body contributions to the nuclear
axial current operator.

Because of the significant progress made with the CHH method in
the description of the four-nucleon bound and continuum
states~\cite{Viv95,Viv98}, the $hep$ reaction has been re-examined in
Refs.~\cite{Mar01,Mar00} and Ref.~\cite{Par03}.  All S- and P-wave 
capture states are calculated, using AV18/UIX interaction
and retaining the $q$-dependence of the nuclear weak transition operator.
In Refs.~\cite{Mar01,Mar00} the TCO model for the weak current is used, 
while in Ref.~\cite{Par03} the EFT model is used, as discussed 
in Sec.~\ref{subsec:wkc}.  The momentum cutoff $\Lambda$ was varied 
between 500 and 800 MeV.  The results for $S(0)$ summarized in
Table~\ref{tab:hepsf} show that the P-wave capture
channels are very important and give about 33\% of the calculated
$S$-factor.  Contributions from the D-wave channels are expected to
be small~\cite{Mar01,Mar00}.  The $\Lambda$ dependence in the $^3S_1$
channel is the result of the large cancellation between the one-body
and the two-body contributions found in Refs.~\cite{Mar01,Sch92,Mar00}; 
such dependence is much smaller in all other channels.

\begin{table}
\caption{Contributions to $S(0)$ for $hep$ from individual partial waves 
in units of $10^{-20}$ keV$\cdot$b.}
\label{tab:hepsf}
\begin{tabular}{ccccccccc}
\hline
 & $\Lambda$ (MeV) & ${}^1S_0$ & ${}^3S_1$ & ${}^3P_0$ & 
   ${}^1P_1$ & ${}^3P_1$ & ${}^3P_2$ & Total \\
\hline
                  & 500 & 0.02 & 7.00 & 0.67 & 0.85 & 0.34 & 1.06 & 9.95 \\
Ref.~\cite{Par03} & 600 & 0.02 & 6.37 & 0.66 & 0.88 & 0.34 & 1.06 & 9.37 \\
                  & 800 & 0.02 & 4.30 & 0.66 & 0.91 & 0.34 & 1.06 & 7.32 \\
Ref.~\cite{Mar01,Mar00} 
                  &     & 0.02 & 6.38 & 0.82 & 1.00 & 0.30 & 0.97 & 9.64 \\
\hline
\end{tabular}
\end{table} 
 
The energy dependence of the $hep$ $S$-factor was found in
Refs.~\cite{Mar01,Mar00} to be rather weak at 0, 5 and 10 keV. In
fact, the calculated $S$-factor at 10 keV, close to the $hep$ solar
Gamow peak, is 10.1$\times10^{-20}$ keV$\cdot$b, only 4\% larger than
that at 0 keV.  In Ref.~\cite{Mar01}, $S(0)$ was also calculated using
the AV18 and the AV14/UVIII interaction models.  The AV14/UVIII (AV18)
result is 4\% (20\%) larger than that for AV18/UIX, demonstrating
the need to use a Hamiltonian that accurately reproduces the
properties of the three- and four-nucleon bound and scattering states.

Following the results of Refs.~\cite{Mar01,Mar00}, the Standard Solar
Model prediction for the $hep$ neutrino flux has been recently 
readjusted to be 9.3 $\times 10^3$ cm$^{-2}$s$^{-1}$~\cite{Bah01}, 
4.4 times larger than the SSM98 prediction, but still about 4 times smaller
than the latest SK result based on LMA~\cite{Fuk02}.

\subsection{The \dalphg\ radiative capture}
\label{sec:A=6}
The reaction \dalphg\ is isoscalar, so direct $E_1$ transitions are
suppressed.  The $M_1$ transition from S-wave initial states is also
suppressed because the ground state is an approximate eigenstate of
the $M_1$ operator.  As a result, the reaction at most energies
between 0.2 and 3 MeV is almost purely $E_2$ in nature, arising from
the $D$ waves.  The cross section is in the nanobarn range
and difficult to measure.  Nonetheless, there are two sets of cross
section measurements \cite{Rob81,Mohr94}, and one set of capture cross
sections inferred from the breakup process $^{208}\mathrm{Pb}+\!
^6\mathrm{Li} \rightarrow\ ^{208}\mathrm{Pb}+ \alpha +d$ \cite{Kie93}.
This small cross section helps preclude significant
production of \nuc{6}{Li} in the big bang.

\begin{figure}
\centerline{\epsfig{file=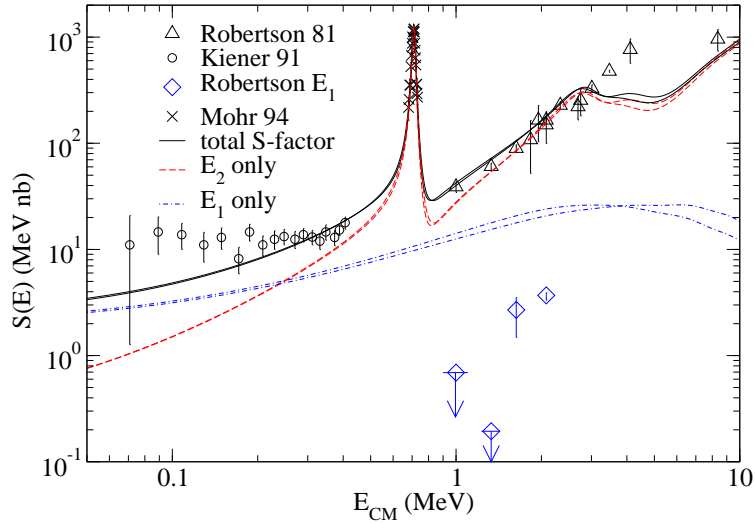,height=10cm,angle=270}}
\caption{$S$-factor for the process \dalphg, showing the variation in
theoretical curves due to choice of $V_{LSJ}$.  See text for a 
discussion of the Robertson $E_1$ \protect\cite{Rob81} points.}
\label{fig:adcapture}
\end{figure}

We show our computed cross sections \cite{NWS01} with the data in
Fig. \ref{fig:adcapture}.  These calculations include all initial
states with $L\leq 3$ and the $M_1,M_2,E_1,E_2,$ and $E_3$ operators.
The two sets of theory curves correspond to two different choices of
$V_{LSJ}$ to generate the inter-cluster correlation $\psi_{\alpha
d}^{LSJ}(r_{\alpha d})$ of Eq.(\ref{eqn:freeschroedinger}).  There is
very little dependence on $V_{LSJ}$, probably because it is
well-constrained in the $D$ waves by low-lying resonances.

The total cross sections provide a good match to the data at the
narrow $3^+$ resonance at 711 keV, partly because $V_{LSJ}$ was
adjusted to produce the location and width of this resonance
accurately~\cite{TUNL} in the scattering phase shifts.  
At energies between the resonance and 3 MeV,
the energy dependence of the cross section provides a good fit to the
Robertson data~\cite{Rob81}.  Beyond 3 MeV, the model fails to
reproduce the data, probably because of omitted couplings to channels
of nonzero isospin.  To match the normalization of the data, our curve
would have to be renormalized by a factor $0.85$, and the residuals of
the data with respect to the renormalized curve are of the order of
10\% even for the higher-quality of the two data sets reported in
Ref. \cite{Rob81}.  We conclude that the theoretical model has come
out quite well.  This probably reflects a
good asymptotic normalization coefficient for projection of the 
\nuc{6}{Li} ground state onto $\alpha d$ cluster states.

In the region below 500 keV, there are only indirect cross sections
(and a rather high upper limit\cite{Cecil}) for comparison, inferred
from the process $^{208}\mathrm{Pb}+\!  ^6\mathrm{Li} \rightarrow\
^{208}\mathrm{Pb}+ \alpha +d$ \cite{Kiener91}.  This energy range is
very interesting, because at about 250 keV, we compute the isoscalar
$E_1$ transition to be half of the total cross section.  It becomes
more important at lower energies, as the centrifugal barrier disfavors
the $E_2$ contributions arising from the $D$ waves.  We do not find the
same energy dependence as the indirect measurements, despite our
inclusion of terms up to third order in the 
long-wavelength approximation (LWA) $E_1$ operator that are
important because of the isoscalar transition.  The energy dependence
of the data suggests S-wave capture, but we compute the cross
section for $E_2$ capture from the $S$ wave to be two orders of
magnitude smaller.  The angular distributions of the breakup data are
consistent with purely $E_2$ transitions, but the technique also gives
much less weight to $E_1$ transitions than to $E_2$.

Although the $E_1$ operator makes a small contribution to the total
cross section above 1 MeV, it interferes with the dominant $E_2$
amplitudes to produce a forward-backward asymmetry in the distribution
of emitted photons.  It is a persistent difficulty of capture models
that they agree with the laboratory data on neither the sign nor the
magnitude of the asymmetry\cite{Rob81,Ryzh95}.  Ours also disagrees
with the data, despite the inclusion of several corrections to the LWA
$E_1$ operator, because the leading effect remains the center-of-energy
effect arising from the difference in mass per nucleon between the two
initial-state nuclei.  We note, though, that the upper limits on
asymmetry at 1.0 and 1.63 MeV are much smaller than one would expect
from extrapolating to lower energies the $E_1$ contribution measured at
2 MeV in essentially any capture model.  This suggests that the
shortcoming may be in the data.

\subsection{The \hag\ and \tag\ radiative captures}
\label{sec:A=7}

\begin{figure}
\centerline{\epsfig{file=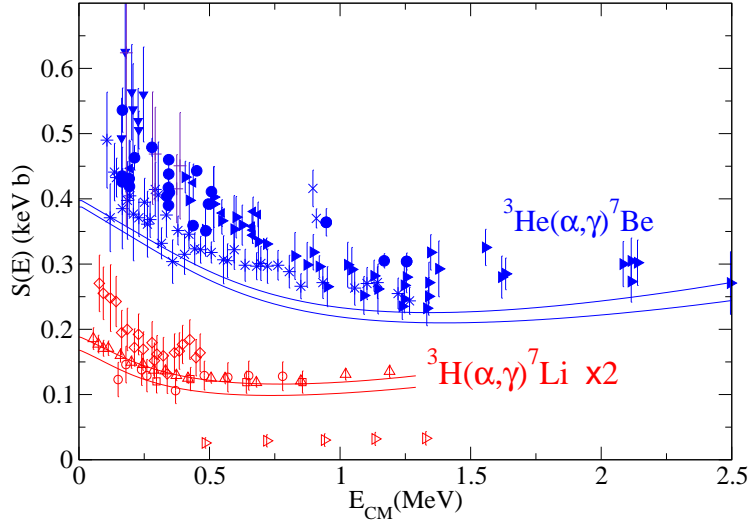,height=10cm,angle=270}}
  \caption{$S$-factors for $\alpha$ captures forming \nuc{7}{Li} and
\nuc{7}{Be}, showing the spread in the computed values due to choice
of $V_{LSJ}$; values for \tag\ are multiplied by 2.}
  \label{fig:mass7capture}
\end{figure}

The $\alpha$-capture processes \hag\ and \tag\ are closely related,
and have also been computed by the VMC method \cite{Nol01}.  
We used the same phenomenological potentials $V_{LSJ}$
of Sec. \ref{sec:freevmc} to describe the continuum in both systems,
except that the charges in the Coulomb terms and the reduced masses
were adjusted to match each system.  Both processes are $E_1$
transitions, originating from $S$ waves at the lowest energies and
also partly from $D$ waves at the higher end of the energy range shown
in Fig. \ref{fig:mass7capture}.  We computed operators for the same
$L$ range and multipolarities as for \dalphg\ and found that no
transitions besides $E_1$ matter above the 0.1\% level for either
reaction.  Corrections to the LWA $E_1$ operator are also negligible
(though the center-of-energy correction, omitted because it is
expensive to compute, should be on the order of 3\% in these cross
sections).  For both reactions, capture may occur into either of two
bound states, with a branching ratio of roughly 0.4.

The computed cross sections \cite{Nol01} for the process \tag\ are
shown and compared with the data
\cite{tag-holmgren,tag-schroeder,tag-griffiths,tag-burzynski} in
Fig. \ref{fig:mass7capture}.  By far the most precise experiment to
measure this cross section is that of Brune et al. \cite{Bru94}
(upward-pointing open triangles in Fig. \ref{fig:mass7capture}).  Our
results for this reaction have a dispersion of about $\pm 5\%$ among
calculations for different choices of cluster-cluster potential
$V_{LSJ}$.  This dispersion, wider than for \dalphg\ or \hag, probably
reflects a substantial contribution to the capture from small
$\alpha$\nuc{3}{H} separations.  Taking account of this dispersion and
of a roughly 10\% normalization uncertainty from our Monte Carlo
integration, we are in excellent agreement with all the modern data.
The ratio of the cross section for capture into the excited state of
\nuc{7}{Li}* to that for the ground state is almost independent of
$V_{LSJ}$.  While the calculation reproduces the approximate absence
of energy dependence seen in the branching ratio data, the predicted
ratio itself is about 15\% below the data.  (The branching data have a
4\% normalization uncertainty.)

The computed energy dependence of the \tag\ total cross section agrees
reasonably well with the Brune data, for a reduced $\chi^2$ of 2.5.
This compares favorably with other calculations in the
literature.  At the highest-energy data points, where the
$S$-factor is increasing due to a growing contribution from D-wave
capture, our calculation has a slightly shallower energy dependence
than the data.  We also find a significantly shallower slope for the
$S$-factor at zero energy than all but one other theoretical
calculation \cite{Mohr93}. 

Fig. \ref{fig:mass7capture} also shows the computed \cite{Nol01} and
measured
\cite{he3ag-hilgemeier,he3ag-kraewinkel,he3ag-nagatani,he3ag-parker,he3ag-robertson,he3ag-osborne}
$S$-factors for the process \hag.  Because of the larger Coulomb
barrier, this reaction is more peripheral than \tag, and this is
probably why its cross section depends less on $V_{LSJ}$.  This is
probably also why there is excellent agreement in the energy
dependence of the cross section both with the laboratory data and with
older theoretical models.  The cross section normalizations of the
various experiments are not completely consistent with each other, but
they all show higher normalizations than do the results of our
calculation, ranging from a 10\% to a 100\% discrepancy relative to
our curve.  While a 10\% discrepancy is within the range allowed by
the combined uncertainties of the experiments, of $V_{LSJ}$, and of
our Monte Carlo integration, the more discrepant data sets are in
clear disagreement with our result.  Since the reaction is peripheral,
and the branching ratio for capture into the two bound states is well
reproduced, this suggests that the wave functions for both \nuc{7}{Be}
bound states have incorrect values for their asymptotic normalizations
in the $L=1$ $\alpha$ $^3\mathrm{He}$ channels, and by about the same
factor.  Because of the discrepancies among experimental data, this
reaction is a case where a microscopic model that can reliably predict
the cross section normalization has the potential to shed light on the
nature of the experimental difficulties and to be very useful to
astrophysics.  Future calculations with wave functions that are more
accurate solutions for the underlying two- and three-nucleon
interactions (variational or GFMC continuum states and GFMC bound
states) should achieve that goal.

\subsection{\nuc{7}{Be} weak decay}
\label{sec:mass7weak}
The $^7$Be nucleus decays by electron ($\epsilon$-)capture to the
ground state of $^7$Li and to its first-excited state at 0.478 MeV.
These decays show up in the solar neutrino spectrum as two sharp lines
at 0.862 and 0.384 MeV, respectively.  The two ground states are both
spin-$\frac{3}{2}$ states, while the $^7$Li* excited state is
spin-$\frac{1}{2}$.  Consequently there are both Fermi and
Gamow-Teller (GT) matrix elements to $^7$Li, but only GT matrix
elements to $^7$Li*.  These have been calculated using both Type I and
Type II VMC wave functions as described in Sec.~\ref{sec:boundvmc}, in
impulse approximation (IA) and with one-body relativistic, two-body
meson-exchange-current, and isobar contributions~\cite{SW02}.

The Fermi matrix element is $F = -\sqrt{2J_f+1}$ for Type I wave
functions due to isospin symmetry, but is slightly less, $F= -1.999$,
for Type II wave functions which include long-range Coulomb
correlations.  The results for GT matrix elements are shown in
Table~\ref{tab:weak}, where we also show the $^3$H matrix elements
used to fix terms beyond IA.  (The $A=7$ IA results have been updated
from Ref.\cite{SW02} to reflect recent improvements in the VMC wave
functions.)  The $^3$H IA matrix elements have been calculated with
both CHH and VMC wave functions with agreement to better than 0.5\%.
The relativistic, mesonic, and isobaric corrections total about 4\% in
$^3$H, and 3\% in $^7$Be.  The two types of VMC wave functions for
$A=7$ give the same results within 1\%.  Compared to experiment,
however, the total GT matrix elements are too small by 7\% for the
ground state and 5\% for the excited state; the corresponding
half-life is calculated to be 62.3 days in IA, and 58.2 days in total,
compared to the experimental 53.22$\pm$0.06 days.  The branching ratio
to these two states is calculated to be 10.20\% in IA, and 10.33\% in
total, compared to an experimental value of 10.44$\pm$.04\%.  These
calculations should be repeated with the more precise GFMC wave
functions, and with an improved Hamiltonian such as AV18/IL2 that does
not underbind the $A=7$ nuclei.

\begin{table}
\caption{Gamov-Teller matrix elements for A=3,7 nuclei.}
\begin{tabular}{lccccccc}
\hline
& \multicolumn{2}{c}{CHH} 
& \multicolumn{2}{c}{VMC(I)} 
& \multicolumn{2}{c}{VMC(II)} & \\
& IA    & Full  & IA       & Full     & IA       & Full     & Expt \\
\hline
$^3$H $\rightarrow ^3$He  
& 1.597 & 1.658 & 1.602    &          &          &          & 1.658 \\
$^7$Be $\rightarrow ^7$Li 
&       &       & 2.345(3) & 2.419(5) & 2.367(2) & 2.433(5) & 2.599 \\
$^7$Be $\rightarrow ^7$Li*
&       &       & 2.142(2) & 2.200(3) & 2.141(1) & 2.205(3) & 2.323 \\
\hline
\end{tabular}
\label{tab:weak}
\end{table}

\section{Conclusion and prospects for the future}
\label{sec:prospects}
Recent advances in the descriptions of nuclear Hamiltonians, wave
functions, and electroweak currents -- as well as computer speed --
may be fruitfully applied to several reaction and decay processes of
interest for astrophysics.  Calculations based on realistic $NN$
interactions and currents may be relied upon for a more fundamental
understanding of the nuclear systems than previous treatments.  In the
case of the weak captures, experimental fitting or verification of
cross sections is not possible, so it is absolutely crucial to have
the most fundamental theoretical understanding possible.  The
radiative captures are in many cases observable in the laboratory, but
calculations are still necessary as tests of wave functions and
currents, as guides for extrapolating laboratory data to lower energies,
and (eventually) as independent sources of information on the absolute
cross section.

There remains considerable opportunity for future efforts in this
area.  For captures producing p-shell nuclei, work is under way to
develop the VMC technique for continuum states so that they can be
derived directly from the underlying $NN$ interaction.  Future
improvements will include the use of the new Illinois $3N$
interactions to generate wave functions, as well as the development of
methods to compute matrix elements from the (essentially exact) GFMC
wave functions.  More direct extraction of the asymptotic
normalization coefficients from the quantum Monte Carlo wave functions
may in the future prove to be very useful for direct captures.

More reactions will also be accessible in the future, partly because
of improvements in algorithms and in computer speed.  Ongoing work
will extend theoretical calculations of the reaction
$d(p,\gamma)^3\mathrm{He}$ to higher energy.  The reaction
$^7\mathrm{Be}(p,\gamma)^8\mathrm{B}$ is an important problem in light
of new laboratory data and the need to extrapolate them to solar
energies.  The isospin mirror process,
\nuc{7}{Li}$(n,\gamma)$\nuc{8}{Li} has been extensively measured
\cite{Tra03} and will provide an interesting test of the method.  Also
in progress is a study of the weak decay of \nuc{8}{B}, which is much
like the $hep$ process in the presence of a spectator
$\alpha$-particle.  Some longer-term prospects are
\nuc{8}{B}$(p,\gamma)$\nuc{9}{C} and \nuc{4}{He}$(\alpha
n,\gamma)$\nuc{9}{Be} reactions that may provide bridges across the
$A=5$ and $A=8$ stability gaps in some astrophysical environments.
The latter will most likely have to be treated as a neutron capture on
the lowest $0^+$ state of \nuc{8}{Be}.

As more is learned about solar neutrinos, the big bang, and other
areas of nuclear astrophysics, the demand for greater precision in
nuclear inputs to astrophysical calculations is increasing.  Ongoing
advances in theoretical treatments of reactions and decays among the
light nuclei will continue to help meet this demand.

We wish to thank our collaborators J. Carlson, A. Kievsky, K. Kubodera,
D.-P. Min, V.R. Pandharipande, T.-S. Park, S.C. Pieper, D.O. Riska, 
M. Rho, S. Rosati, and M. Viviani, 
for their many contributions to the present subject.
The work of R.S. was supported by the U.S. DOE Contract 
No. DE-AC05-84ER40150, under which the Southeastern Universities 
Research Association (SURA) operates the Thomas Jefferson National 
Accelarator Facility, and that of K.M.N. and R.B.W. by the DOE, 
Nuclear Physics Division, under Contract No. W-31-109-ENG-38.

\bibliographystyle{elsart-num}
\bibliography{final}

\end{document}